\documentclass[conference, letterpaper, 10pt]{IEEEtran}
\ifCLASSINFOpdf
\else
\fi
\usepackage{amsmath}
\usepackage{algorithmic}
\usepackage{algorithm}
\usepackage{graphicx}
\usepackage{amssymb}
\usepackage{epstopdf}
\usepackage[flushleft]{threeparttable}
\usepackage[keeplastbox]{flushend}
\usepackage{balance}
\usepackage{eso-pic}
\newcommand\blfootnote[1]{%
  \begingroup
  \renewcommand\thefootnote{}\footnote{#1}%
  \addtocounter{footnote}{-1}%
  \endgroup
}
\begin{document}
\bstctlcite{IEEEexample:BSTcontrol}
\title{LTE PHY Layer Vulnerability Analysis and Testing Using Open-Source SDR Tools}
\author{\IEEEauthorblockN{Raghunandan M. Rao\IEEEauthorrefmark{1}, Sean Ha\IEEEauthorrefmark{1}, Vuk Marojevic\IEEEauthorrefmark{1},
Jeffrey H. Reed\IEEEauthorrefmark{1} 
}
\IEEEauthorblockA{\IEEEauthorrefmark{1}Bradley Department of Electrical and Computer Engineering\\ Virginia Tech, Blacksburg, Virginia, USA\\ 
Email: \{raghumr, seanha65, maroje, reedjh\}@vt.edu}
}

\maketitle
\begin{abstract}
This paper provides a methodology to study the PHY layer vulnerability of wireless protocols in hostile radio environments. Our approach is based on testing the vulnerabilities of a system by analyzing the individual subsystems. By targeting an individual subsystem or a combination of subsystems at a time, we can infer the weakest part and revise it to improve the overall system performance. We apply our methodology to 4G LTE downlink by considering each control channel as a subsystem. We also develop open-source software enabling research and education using software-defined radios. We present experimental results with open-source LTE systems and shows how the different subsystems behave under targeted interference. The analysis for the LTE downlink shows that the synchronization signals (PSS/SSS) are very resilient to interference, whereas the downlink pilots or Cell-Specific Reference signals (CRS) are the most susceptible to a synchronized protocol-aware interferer. We also analyze the severity of control channel attacks for different LTE configurations. Our methodology and tools allow rapid evaluation of the PHY layer reliability in harsh signaling environments, which is an asset to improve current standards and develop new and robust wireless protocols.

\textit{Index Terms} -- Long-term evolution (LTE); Methodology; Protocol-aware Jamming;
Software-Defined Radio (SDR); Open source tools.
\end{abstract}

\IEEEpeerreviewmaketitle

\blfootnote{This is the author's version of the work. For citation purposes, the definitive version of record of this work is: R.M. Rao, S. Ha, V. Marojevic, J.H. Reed, ``LTE PHY Layer Vulnerability Analysis and Testing Using Open-Source SDR Tools'', IEEE MILCOM 2017, 23-25 Oct. 2017.}

\section{Introduction}\label{Sec1_Intro}
Wireless communications infrastructure is a vital resource to the economy and defense of the nation, and its protection is becoming increasingly important. While the commercial sector continues to increase the use of cellular networks for Internet access, public safety networks and the military plan to leverage commercial 4G cellular technology for their communication needs in emergency situations. The Internet of Things (IoT) will also impact the number of devices connecting to the cellular network. Even though the architecture of 5G is still an open question, there will be key differences with regards to many characteristics of 4G technology such as the ability to flexibly use spectrum; order of magnitude reduction in latency; and most importantly, improved security and privacy \cite{YangWang_5GSec_2015}.

Wireless communication networks provide a multitude of critical and commercial services and serve an ever increasing number of users. 
Ultra-reliable and secure wireless networks are necessary for national security and public safety, in order to provide efficient situation assessment, time-critical assistance and swift response to potential threats. This is also crucial for autonomous vehicular networks and unmanned aerial systems, where even a partial breakdown can have disastrous consequences. 

Security and privacy has been a recurrent problem in wireless communications since the days of analog 1G cellular networks. Each cellular generation has undergone its share of research and development related to wireless network attacks and countermeasures because each new feature introduced can have a potential vulnerability that can be exploited by an attacker. 
Attacks to wireless networks have been the topic of research for several years \cite{Pelechrinis_DoS_2011}-\cite{Zou_Hanzo_WiSec_2016}. 

Lazos et al. \cite{Lazos_adhoc_Cont_2009} propose a randomized distributed channel establishment scheme based on frequency hopping to address the problem of control channel jamming in ad-hoc networks. Bicakci et al. \cite{Bicakci_DoS_80211_2009} focus on combating Denial of Service (DoS) attacks in 802.11 devices using practical hardware, software and firmware solutions. Liu et al. \cite{Liu_BroadcastJam_2011} develop a time-delayed broadcast scheme that partitions the broadcast operation into a series of unicast transmissions for spread spectrum systems with insider threats. Chiang et al. \cite{Chiang_CroLayJam_2011} introduce a code-tree system for circumventing cross-layer jamming attacks in wireless broadcast networks. Hidden Markov Models (HMMs) are adopted in \cite{He_ByzantAtt_Def_2013} to characterize the sensing behaviors of legitimate and malicious users in collaborative spectrum sensing settings.

LTE is prone to protocol-aware interference, since the standards documentation is openly available. There has been prior work related to RF Jamming of LTE signals. Kakar et al. \cite{Jaber_Marc_PCFICH_2014} investigate
the performance of the Physical Control Format Indicator Channel (PCFICH) under harsh wireless
conditions and propose strategies to mitigate interference on the PCFICH. Lichtman et al. \cite{Lichtman_PUCCH_2014}
consider the problem of targeted interference on the Physical Uplink Control Channel (PUCCH) and propose detection and mitigation strategies to counter protocol-aware jammers. Labib et al. \cite{Labib_LTE_2015} introduce and demonstrate \emph{LTE control channel spoofing}, which refers to spoofing by a fake eNodeB by transmission of a partial LTE downlink frame. DoS was found to be the result of transmission of the partial LTE downlink frames containing only the fake control channels, at a relatively higher power level w.r.t. the legitimate eNodeB. The authors also propose mitigation strategies that required simple modification to the cell selection process of LTE.
In \cite{Lichtman_Jam_2016}, a comprehensive threat assessment of LTE/LTE-A is provided, highlighting the vulnerabilities of various LTE physical channels and signals. A survey of mitigation techniques against various jamming/spoofing attacks is also provided. Marojevic et al. \cite{Vuk_VTCFall2017} provide empirical results for jamming of LTE control channels in a production LTE system, and propose jammer detection methods using performance counters and key performance indicators (KPIs).

This paper provides a methodology for assessing the PHY layer vulnerabilities of cellular networks. Our analysis uses LTE as an example, but the methodology applies to many other present and future protocols. The contribution of paper is threefold. We provide 
\begin{enumerate}
\item a methodology for PHY layer testing of wireless protocols, 
\item open-source software enabling research and education using software-defined radios (SDRs), and 
\item empirical results with open-source SDR-based 4G LTE systems. 
\end{enumerate}
The rest of the paper is organized as follows: Section \ref{Sec2_Bkgrnd} reviews the LTE downlink control signaling which is the focus of our analysis in this paper. Section \ref{Sec3_TestMeth} introduces our methodology, Section \ref{Sec4_ResAna} provides experimental results and analysis for LTE and section \ref{Conc} concludes the paper.

\section{Control Channels in the LTE Downlink}\label{Sec2_Bkgrnd}
Control channels in wireless networks are essential for providing the capability for the system to operate properly and efficiently. Using LTE as an example below we briefly review some of the important Downlink (DL) LTE physical control channels that are relevant for the experimental analysis discussed in this work. Note that in LTE the Base Station is referred to as the evolved NodeB (eNB) and the user terminal is referred to as the user equipment (UE).
\subsubsection*{Primary and Secondary Synchronization Signals (PSS/SSS)}
LTE uses two synchronization signals: the Primary Synchronization Signal (PSS) and Secondary Synchronization Signal (SSS).  The PSS and SSS are broadcast by the eNB for slot and frame synchronization.  The PSS is constructed from Zadoff-Chu sequences due to its strong Constant Amplitude Zero Autocorrelation (CAZAC) properties. 

The SSS is constructed from two maximal length sequences (M-sequences).  As all communication between the UE and eNB requires time synchronization, disrupting the PSS or SSS will not cause an immediate Denial of Service (DoS), but will instead prevent a) new UEs from accessing the cell, and b) idle UEs from re-synchronizing with the cell.

\subsubsection*{Physical Broadcast Channel (PBCH)}
The PBCH contains the Master Information Block (MIB) which informs the UE about the downlink bandwidth, resource length of the Hybrid ARQ (HARQ) Indicator Channel (PHICH), and the System Frame Number (SFN) for frame synchronization. The PBCH is mapped to the central 72 subcarriers of four consecutive OFDM symbols per frame. It is QPSK modulated with a 16-bit CRC, but with an effective coding rate of $1/48$.

\subsubsection*{Physical Downlink Control Channel (PDCCH)}
The PDCCH carries critical control information, such as UE resource allocation, modulation and coding scheme (MCS) of user data, information about the HARQ parameters and precoding matrices for MIMO. It is QPSK modulated with rate-$1/3$  convolutional coding and occupies the first few OFDM symbols of each subframe. During the initial cell access procedure, it informs the UE of the first System Information Block (SIB1), without which the UE will be unable to complete the cell attachment process. Additionally, after cell attachment, it would be impossible for the UE to obtain service and decode its data if the PDCCH is improperly decoded.

\subsubsection*{Physical Control Format Indicator Channel (PCFICH)}
The PCFICH contains information regarding the size of the PDCCH and is sent at the beginning of each subframe. It contains the Control Format Indicator (CFI) which is 2 bits in length, and is encoded using a code rate of $1/16$. Therefore corrupting the PCFICH will also corrupt detection of the PDCCH at the UE.

\subsubsection*{Cell-Specific Reference Signals (CRS)}
The CRS carries downlink pilot symbols that are used for channel estimation, quality assessment and equalization. CRS is QPSK-modulated and use a Gold sequence of length 31, which is initialized using the cell ID value. The CRS symbols are distributed sparsely in time and frequency, occupying about 5\% of the Resource Elements (REs). The first position $k_0$ of the CRS in the LTE DL grid is determined by the Physical Cell Identity (PCI) $N_{c,ID} \in \{0,1,\cdots ,503\}$, by the relation $k_0=N_{c,ID} \text{ mod } 6$ (assuming the range of $k_0 \in \{0,1, \cdots, 5\}$). 



\section{PHY Layer Vulnerability Assessment Methodology and Tools}\label{Sec3_TestMeth}
The methodology that we develop is based on testing the vulnerabilities of a system by analyzing the individual subsystems. By targeting a specific subsystem or a specific combination of subsystems at a time, we can infer the weakest component and revise it to improve the overall system robustness. Hence, we propose tools for assessing the individual system components and metrics for evaluation. Here, we regard the LTE control channel(s) to be the subsystem(s), for the purpose of analysis. Although we present the tools and metrics for LTE, rather than in an abstract way, the concepts are applicable for other wireless protocols as well. The software tools have been presented in \cite{Vuk_VTCFall2017} and are included here for completeness.
\subsection{Open-Source Software Tools}
\begin{figure}[!t]
\centering
\includegraphics[width=3.2in]{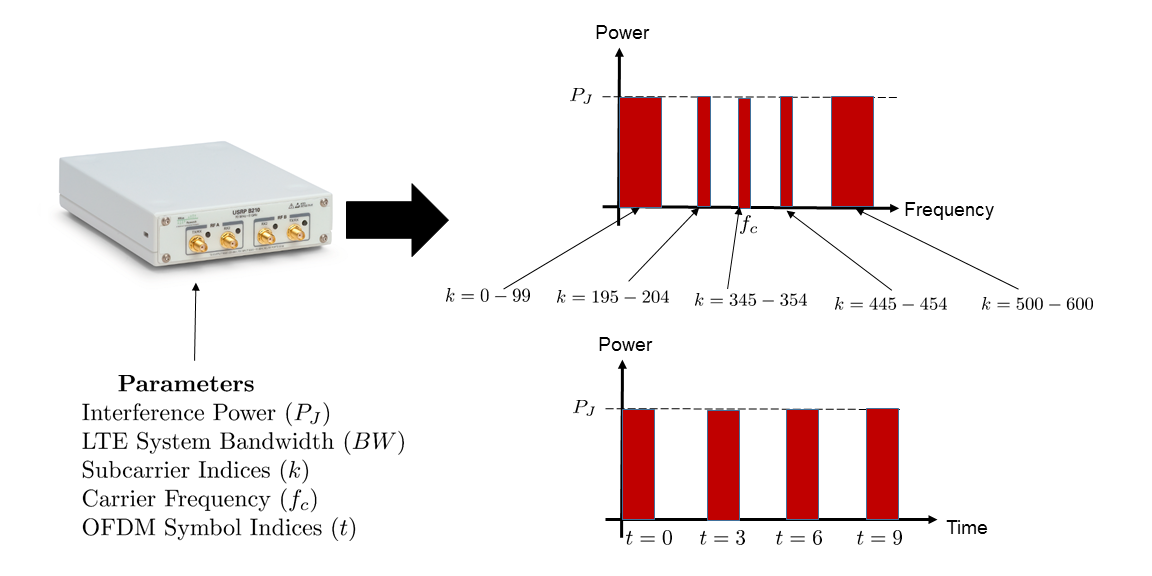}
\caption{OFDM-based parametric interference generation.}
\label{Fig1_JamWF_Gen}
\end{figure}

We propose a parametric framework for interference generation by using the same waveform as the target system. This ensures that there is high correlation in the protocol signaling between the interference and the target waveform(s). In the case of LTE, individual subcarriers and OFDM symbols can be allocated/blanked to rapidly generate wideband, narrowband, and protocol-aware interference over any portion of the DL LTE time-frequency grid. Note that other wireless standards, including IEEE 802.11xx and emerging IoT standards, use OFDM. Fig. \ref{Fig1_JamWF_Gen} illustrates the general framework for OFDM-based parametric interference generation. It allows generating asynchronous and synchronous interference waveforms that we build from the open source LTE library srsLTE \cite{srsLTE} (formerly known as libLTE). The srsLTE library implements the LTE uplink and downlink waveforms and supports commercial off-the shelf hardware such as Universal Software Radio Peripherals (USRPs). Broadly, we classify the interference waveforms based on the time-synchronization requirements of the jammer w.r.t. the target:
\subsubsection{Asynchronous Interference Waveforms}
The asynchronous interference waveform generates interference on specific subcarriers. This type of interference can be of certain duration/duty cycle or continuous or discontinuous in time. We can use this setup to generate any interference to LTE that does not need accurate time alignment with the LTE frame. 
We can also generate a fake PSS and/or SSS signal (for example, to execute PSS/SSS spoofing as in \cite{Labib_LTE_2015}) by replacing OFDM symbols with LTE synchronization sequences. As an example, the spectrum of a 1.4 MHz interference waveform with three blocks of active subcarriers is shown in Fig. 2. Assuming the subcarrier indices $k \in \{0,1, \cdots,71 \}$, the jammer transmits only on subcarriers $k_j$ such that $k_j \in \{0,1, \cdots,35\}$ (540 kHz wide), $k_j \in \{49,50, \cdots, 59\}$ (165 kHz wide) and $k_j \in \{71\}$ (15 kHz wide). 
\subsubsection{Synchronous Interference Waveforms}
Transmitting on top of specific physical channels requires synchronization with the network to determine where the physical channels exist in time. Therefore, we design an interferer that (1) acts as a receiver and synchronizes with the eNB, in this case by acquiring the legitimate PSS and SSS of the LTE cell and (2) synchronously transmits its interference waveform to the target. A configurable timing offset can be specified to account for transmission and other delays. Fig. \ref{Fig3_SyncInt_WF} illustrates the synchronous interference waveform which targets the PSS/SSS.
\begin{figure}[!t]
\centering
\includegraphics[width=3.2in]{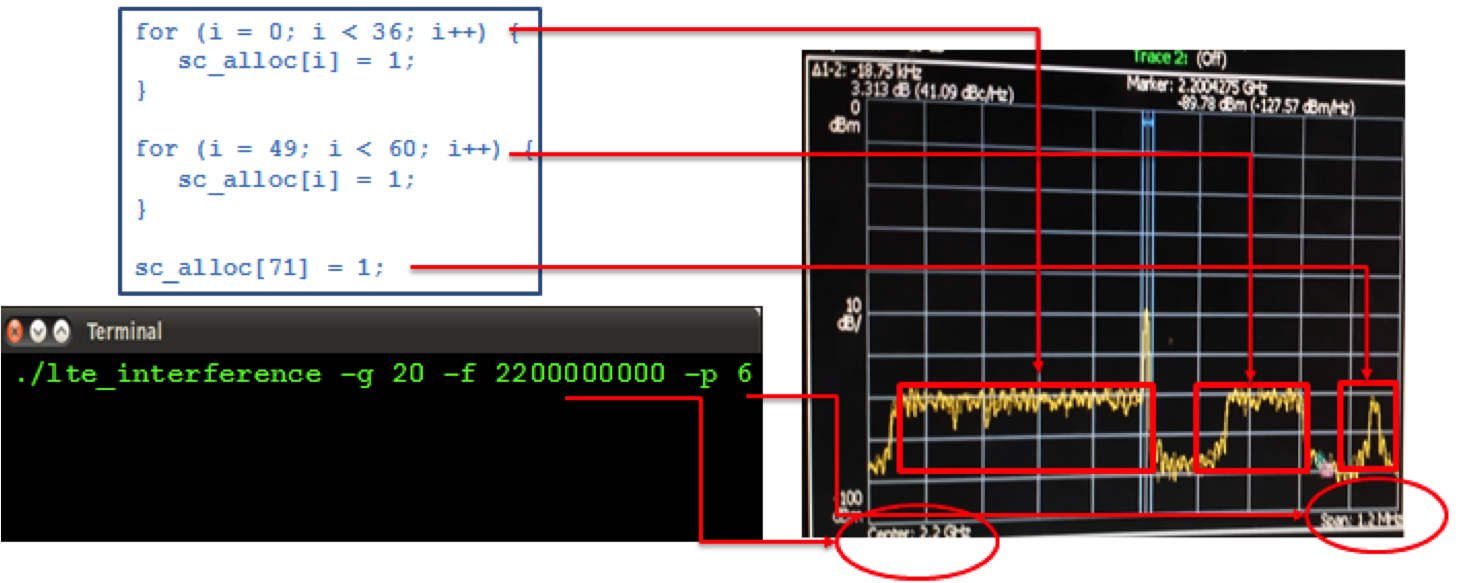}
\caption{Asynchronous interference waveform generation.}
\label{Fig2_liblte_snap}
\end{figure}
\begin{figure}[!t]
\centering
\includegraphics[width=3.2in]{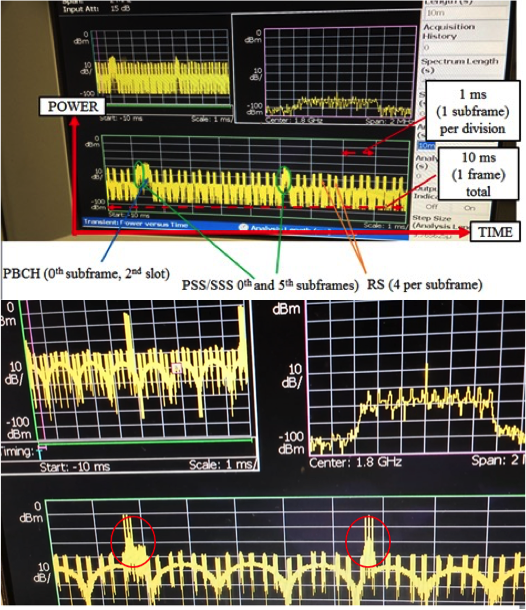}
\caption{Synchronous interference waveform: eNB signals which, for illustration purposes, consist of the PSS/SSS, the PBCH, and the CRS only (top and bottom) and synchronous PSS/SSS interference (bottom).}
\label{Fig3_SyncInt_WF}
\end{figure}
\subsection{Metrics for Performance Evaluation}
We need a uniform metric that captures the RF differences among the various physical control channels and signals. In this regard, we devise a metric based on (a) jammer to signal power ratio (JSR) values, (b) control channel occupancy fraction in the LTE Frame, and (c) relative power w.r.t. the data channel. Equivalent metrics apply for other wireless standards and additional metrics may need to be specified for the specific analysis. For LTE, we define the following quantities: 
\begin{enumerate}
\item Jammer to Signal Ratio per Resource Element $(JSR_{RE})$, 
\item Jammer to Signal Ratio per Frame $(JSR_F)$, and 
\item Relative power w.r.t. the Physical Downlink Shared Channel (PDSCH) $(\rho_{PDSCH})$.
\end{enumerate}
\subsubsection{$JSR_{RE}$}
It is defined as the ratio of the jammer power to that of the LTE signal, assuming that all the LTE resource elements (REs) have the same transmit power. We also consider that the jammer allocates equal power on all subcarriers at all times.

\subsubsection{$JSR_F$}
In our experiments, the jammer is targeting specific control channels, all of which occupy different fractions of the total number of REs per LTE DL frame. To account for this, we define $JSR_F$ given by 
\begin{equation}
JSR_F = JSR_{RE} \times \frac{N_{T,F}}{N_{tot,F}},
\end{equation}
\noindent where $N_{T,F}$ denotes the number of REs allocated to the control channel per LTE frame, and $N_{tot,F}$ the total number of REs per LTE frame.

\subsubsection{$\rho_{PDSCH}$}
LTE allocates different power levels to different physical channels. We introduce $\rho_{PDSCH}$ and define it as $\rho_{PDSCH} = \frac{P_T}{P_{PDSCH}}$, where $P_T$ is the power allocated by the eNB to the control channel, and  
$P_{PDSCH}$ the power allocated to the PDSCH in the same frame. 

These three factors determine the overall jammer to signal ratio as
\begin{equation}
\label{JSR_overall}
JSR_N = JSR_{F} \times \rho_{PDSCH}. 
\end{equation}
The above relation is adapted from \cite{Licht_host_int_LTE_2013} to account for the non-uniformity in power allocation for heterogeneous power allocations across LTE's physical downlink channels and signals. Since $JSR_N$ represents the overall power requirements of the jammer w.r.t. that of a DL LTE signal with unit power per resource element, we use it to compare the relative robustness of the control channels.

In order to compare the robustness of each control channel to targeted interference, we also need to define events at the receiver that constitute a successful jamming attack. Related to that, we also set thresholds for the \emph{probability of error} or \emph{error rate} $P_{err}$, beyond which the control channel is said to be irreversibly corrupted. In essence, such a scenario will lead to the UE operating with corrupted control information that can cause severe system level issues, and in the worst case, DoS. Table \ref{Tab2_JamInd} describes the event  (i.e. error flag in srsLTE) that indicates a successful LTE DL jamming attack.

\begin{table}[!t]
\renewcommand{\arraystretch}{1.1}
\caption{Indicators of all considered jamming strategies. }
\label{Tab2_JamInd}
\centering
\begin{tabular}{|l|c|l|}
\hline
Target & Periodicity & Error Flag \\
\hline
Barrage & Subframe & PDCCH error/PCFICH error/synch error/ \\
& & PBCH error \\
\hline
PSS/SSS & Frame & If frame synchronized: sync loss. If frame \\
& & does not synchronize: Unable to sync \\
\hline
PDCCH & Subframe & $\text{Decoded RNTI}\neq \text{SENTVALUE}$ \\
\hline
PBCH & Frame & $\text{Decoded MIB}\neq \text{Default configuration}$ \\
\hline
PCFICH & Subframe & $\text{Decoded CFI} \neq 2$ \\ 
\hline
CRS & Subframe & PDCCH error/PCFICH error/synch error/\\
& & PBCH error\\
\hline
\end{tabular}
\end{table}

\section{LTE Experimental Results and Analyses}\label{Sec4_ResAna}
\subsection{Experimental Setup using Open-Source SDR Tools}
\begin{figure}[!t]
\centering
\includegraphics[width=3.2in]{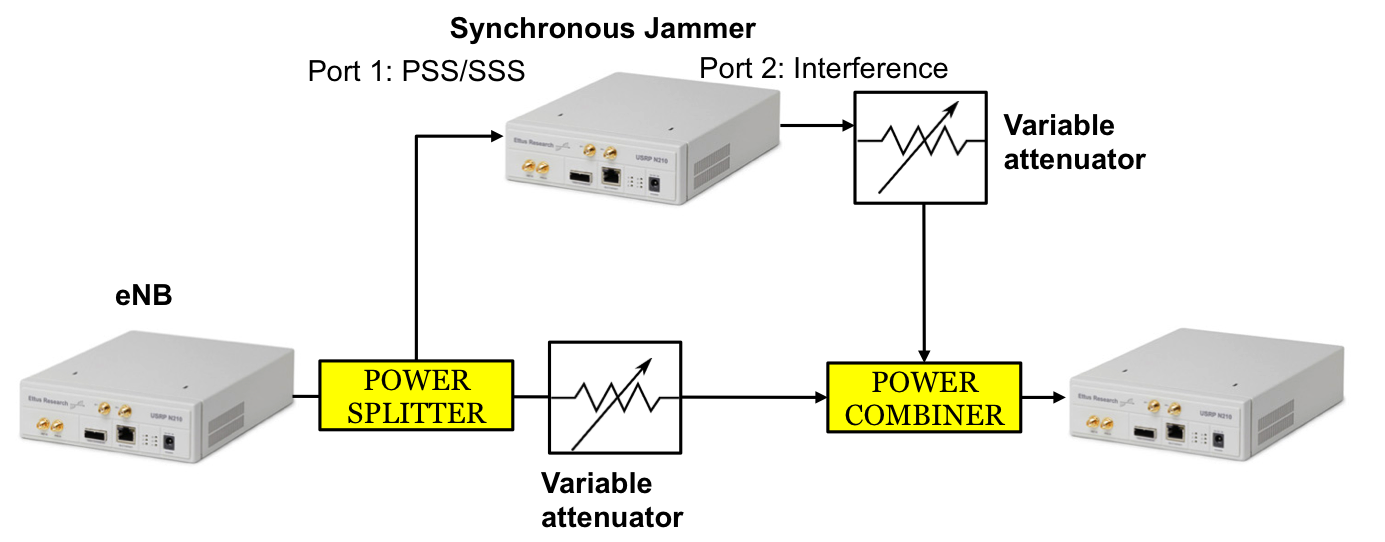}
\caption{Experimental setup of the jamming experiments.}
\label{Fig4_JamExpt_setup}
\end{figure}

Fig. \ref{Fig4_JamExpt_setup} shows the experimental setup using components from Virginia Tech's LTE Testbed \cite{Vuk_LTETest_2017}. It features three Universal Software Radio Peripherals (USRPs) to emulate the LTE eNB, UE and the protocol-aware jammer. The jammer acquires the PSS/SSS from the eNB in order to synchronize the jamming signal with the UE as needed. The LTE DL signal and the synchronous jamming signal are combined using radio frequency (RF) power combiners. Variable attenuators in each RF path provide a means to control the JSR. RF cables are used to connect the components. Therefore, the resulting LTE system performance is interference-limited.

Each considered DL physical control channel is tested using $N_{trial}=10^5$ trials of targeted interference. Each trial composed of transmission of 1 subframe (for PCFICH, PDCCH, CRS and Barrage jamming schemes) and 1 frame (for PSS/SSS and PBCH jamming schemes). Table \ref{Tab2_JamInd} indicates the error event for each jamming scenario. The error rate for each channel $P_{err}$ is defined as the ratio of the the number of frames (or subframes) for which the error flag (defined in Table \ref{Tab2_JamInd}) is active $(N_{err})$, to the total number of frames (or subframes) transmitted $(N_{trial})$. The error rate is calculated using the relation $P_{err} = \frac{N_{err}}{N_{trial}}$. The JSR is computed using (\ref{JSR_overall}), where the term $\rho_{PDSCH}$ is experimentally obtained for each jamming strategy, and is tabulated in Table \ref{Tab1_rhoPDSCH}. 

We define the error threshold $P_{err,th}$ to be the value beyond which DoS occurs. The corresponding values are summarized in Table \ref{Tab3_LTE_Chan_Occupancy}. The choice of $P_{err,th}$ for each strategy can be justified as follows:
\begin{enumerate}
\item PBCH has a very low coding rate of $1/48$, and is spread across 4 OFDM frames. Hence, it is robust against interference and we set a high error threshold of $P_{err,th}=0.9$.
\item PSS/SSS is comprised of interference-resilient signals with very strong correlation properties \cite{Licht_host_int_LTE_2013}. Hence we set an error threshold value of $P_{err,th}=0.5$.
\item PCFICH carries the locations of the PDCCH, which has a fairly high code-rate of $1/3$. Hence, an error threshold of $P_{err,th}=0.1$ is considered to be sufficient to disrupt these channels. 
\item CRS is used for channel estimation and equalization which is crucial to decode all the other control channels. The aggregate $P_{err,th}$ cannot be greater than that of all other critical DL control channels and hence, we set $P_{err,th} = 0.1$. The same reasoning applies for barrage jamming as well.
\end{enumerate}
The overall jammer to signal noise ratio ($JSR_{N,DoS}$) represents the JSR value that results in $P_{err} \geq P_{err,th}$, which in turn results in DoS.

\begin{table}[!t]
\renewcommand{\arraystretch}{1.1}
\caption{Empirically measured values of $\rho_{PDSCH}$.}
\label{Tab1_rhoPDSCH}
\centering
\begin{tabular}{|l|c|}
\hline
Jamming Strategy & $\rho_{PDSCH}$ (dB)\\
\hline
Barrage & 0 \\
\hline
PSS/SSS & -5 \\
\hline
PDCCH & -5\\
\hline
PBCH & -2\\
\hline
PCFICH & -8\\ 
\hline
CRS & -10\\
\hline
\end{tabular}
\end{table}

\subsection{Analysis of DL Jamming Results} \label{Open_SDR_Results}
\begin{figure}[!t]
\centering
\includegraphics[width=3.2in]{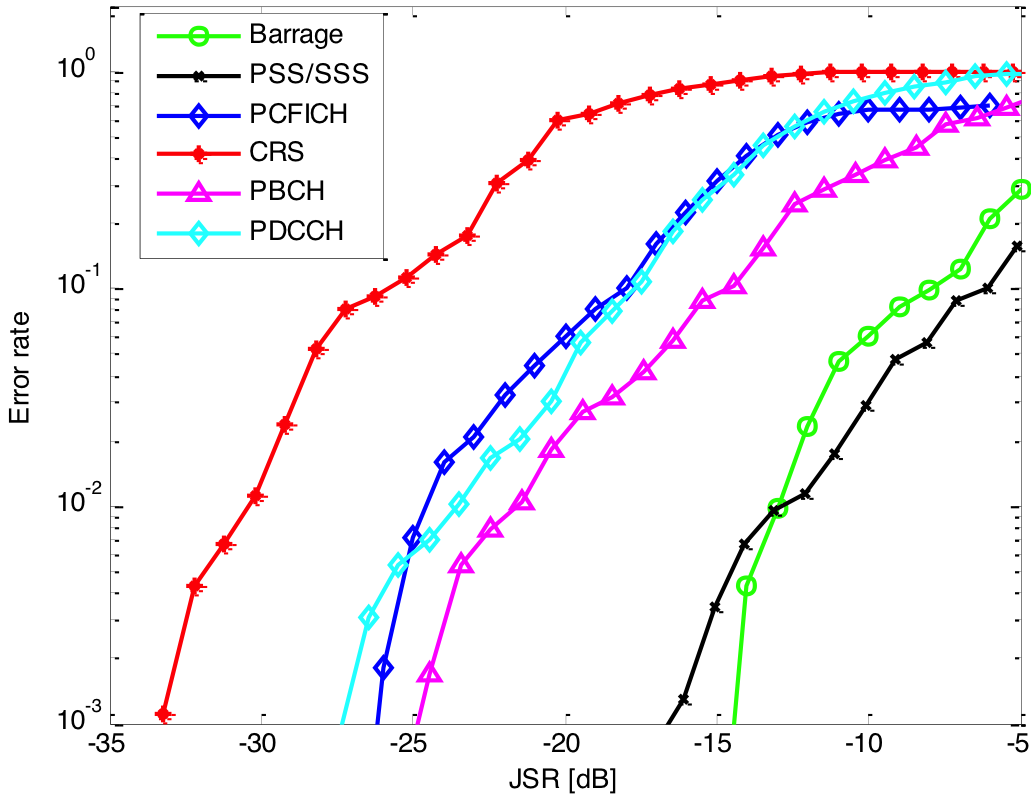}
\caption{Error rate as a function of JSR for the jamming experiments using open-source SDR tools for a 1.4 MHz FD LTE signal.}
\label{Fig5_libLTEJamResults}
\end{figure}

The jamming experiments were performed for all six downlink jamming strategies of Tables \ref{Tab2_JamInd}-\ref{Tab3_LTE_Chan_Occupancy} and the results are shown in Fig. \ref{Fig5_libLTEJamResults}. The necessary JSR to sufficiently corrupt each physical channel and cause DoS, is summarized in Table \ref{Tab3_LTE_Chan_Occupancy}. We qualitatively comment on the complexity of the jamming attack for a 1.4 MHz FD LTE system, based on the degree of synchronization required to jam the subsystem and the contiguity of targeted REs comprising the subsystem. We see that PSS/SSS is the least vulnerable DL physical channel. This is understandable since both the PSS and SSS use very strong correlation sequences (Zadoff-Chu and M-sequence respectively) that are inherently robust against noise and interference. Even though PSS and SSS are very sparse, the jammer needs more power than the LTE DL signal to corrupt them to the point of DoS (due to synchronization loss/inability to synchronize).

\begin{table}[!t]
\renewcommand{\arraystretch}{1.1}
\caption{LTE DL Physical Channel Vulnerabilities for 1.4 MHz FD-LTE.}
\label{Tab3_LTE_Chan_Occupancy}
\centering
\begin{tabular}{|l|c|c|c|c|}
\hline
Target & Fraction  & Complexity & $P_{err,th}$ to & $JSR_{N, DoS}$  \\
& of REs & & cause DoS & (dB)\\
\hline
Barrage & 100 \% & Very Low & 0.1 & -10 \\
\hline
PSS/SSS & 1.23 \% & Medium & 0.5 & 5 \\
\hline
PDCCH & 23.4 \% & Medium & 0.1 & -16 \\
\hline
PBCH & 2.38 \%  & Low & 0.9 & -3 \\
\hline
PCFICH & 0.2 \% & High & 0.1 & -19 \\ 
\hline
CRS & 4.76 \% & High & 0.1 & -26 \\
\hline
\end{tabular}
\end{table}
The PBCH is also resilient to jamming, since (a) it provides only basic initial information, and (b) it has a very low coding rate (cf. Section \ref{Sec2_Bkgrnd}). The PCFICH is robust to interference because of strong encoding, but its sparsity makes it susceptible to jamming relative to PDCCH. However, since jamming the PCFICH has the same outcome as jamming the PDCCH, the 3 dB gain of PCFICH w.r.t. PDCCH jamming is deemed impractical for the jammer since the PCFICH's sparsity makes the jamming attack more complex and countermeasures have been proposed in the literature \cite{Jaber_Marc_PCFICH_2014}.

We see that CRS, in the case of a perfectly synchronized jamming attack, is highly vulnerable. This is due to (a) the sparsity of the CRS in terms of the fraction of REs per OFDM frame and (b) the dependence of all control and data channels on accurate channel estimates. 
In practice, perfect synchronization is difficult to achieve for a sparse and non-contiguously distributed signal such as the CRS. However, the jammer can target CRS subcarriers instead of CRS REs, which is a very simple strategy since the CRS subcarrier locations depend on the cell ID $N_{c,ID}$ and remain constant for all users of the cell. For this more practical case the jammer need not synchronize with its target, thus reducing the complexity of the attack at the cost of a higher JSR.

It is clear that the PDCCH should have the highest priority for anti-jamming countermeasures because the PDCCH can be easily targeted with a relatively low JSR. 
CRS follows close behind, having by far the lowest required JSR to corrupt, but a much higher complexity. 
Similarly, the PBCH and, especially PSS/SSS are not high priority for anti-jamming; while an attack may theoretically cause complete DoS by targeting these subystems, the attackers require nearly as much power as barrage jamming, making them less practical from an attacker's perspective.
\begin{figure}[!t]
\centering
\includegraphics[width=3.2in]{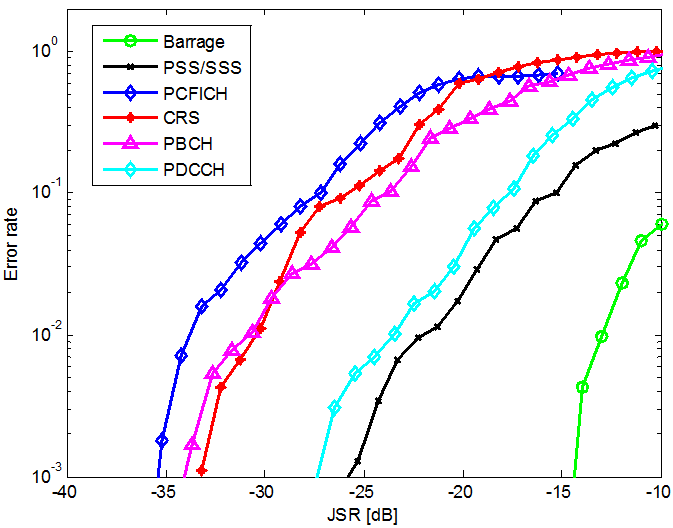}
\caption{Error rate as a function of JSR for the jamming experiments using open-source SDR tools for a 10 MHz FD LTE signal.}
\label{Fig9_libLTEJam_10MHz}
\end{figure}

\subsection{Vulnerability Analysis for Higher LTE Bandwidths}
The results discussed so far apply only to a LTE bandwidth of 1.4 MHz. For higher LTE bandwidths, it is important to note that the following control channels have the same number of resource elements per frame for all LTE bandwidths, (a) PSS/SSS, (b) PBCH and (c) PCFICH. Hence, these control channels can be targeted by the jammer by transmitting the same sum power as for a 1.4 MHz LTE system. For cases (a)-(c) the required JSR ($JSR_{N,DoS}^{(BW)}$) to cause DoS for a LTE system bandwidth $BW$ can then be written as
\begin{equation}
\label{JSR_BW}
JSR_{N,DoS}^{(BW)} = JSR_{N,DoS}^{(1.4\text{ MHz})} - 10\log_{10} \Big(\frac{BW}{1.4 \text{ MHz}} \Big).
\end{equation}

However CRS, PDCCH and barrage jammers need to occupy the entire bandwidth. The $JSR_N$ requirements for all other strategies are a function of the LTE bandwidth. Fig. \ref{Fig9_libLTEJam_10MHz} shows the error rate as a function of the JSR for a 10 MHz FD-LTE system. 

\begin{figure}[t]
\centering
\includegraphics[width=3.2in]{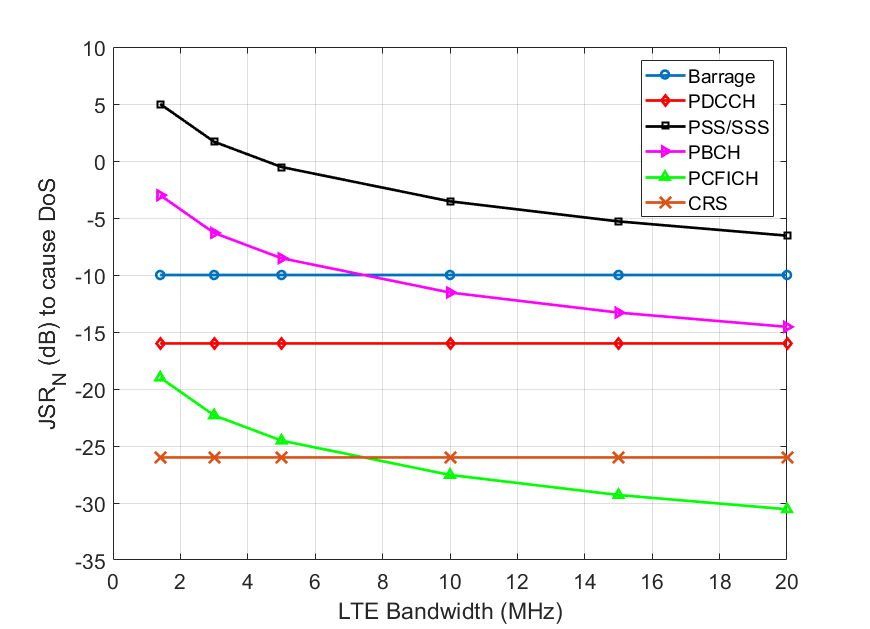}
\caption{$JSR_N$ required versus LTE bandwidth to cause Denial of Service for all considered control channel attacks.}
\label{Fig_JSR_N_DOSvsBW}
\end{figure}

Fig. \ref{Fig_JSR_N_DOSvsBW} shows the variation of $JSR_{N,DoS}$  as a function of LTE system bandwidth ($BW$) for each jamming strategy. We see that (a) PSS/SSS will still be an inefficient attack for system bandwidths of up to 20 MHz, (b) PBCH jamming is more of an issue when compared with barrage jamming only for $BW \geq 10 \text{ MHz}$, (c) PCFICH jamming performs better than CRS jamming for $BW \geq 10 \text{ MHz}$, and (d) CRS jamming always outperforms PDCCH jamming. An obvious way to increase the resilience of the system would be to allocate all the control channel REs (except the PSS/SSS) over the entire LTE bandwidth. This would ensure that the jammer would have to resort to wideband signaling to disrupt the LTE control channels.
\section{Conclusions}\label{Conc}
This paper has introduced a methodology for analyzing the PHY layer vulnerability of wireless protocols. We have developed a software suite for introducing protocol-aware interference to a wireless system. Our software is applicable to OFDM-based protocols. We applied it to target the LTE RF downlink and showed how different subsystems of LTE react differently to protocol-aware interference. The results showed that the LTE synchronization signals (PSS/SSS) are most resilient to synchronized jamming causing DoS. We also observed that CRS, PCFICH and PDCCH are the most vulnerable to protocol-aware jamming. The results presented in this paper are based on open-source SDR-based LTE systems. Other systems will need different metrics and software for testing, but the proposed methodology can be applied for (1) methodical testing, (2) improving the robustness of the subsystem and hence the entire system as in \cite{Labib_ComMag_2017} and (3) designing control channels to achieve a desired level of robustness.  

As LTE evolves, these vulnerabilities will need to be taken into consideration while applying LTE for military and public safety networks, and for designing interference-resilient next-generation wireless protocols. Tradeoffs will be necessary to balance efficiency, system complexity and robustness to smart attackers. This will continue to be an important research area that will significantly contribute to protocols and procedures of 5G wireless networks.

\section*{Acknowledgements}
The authors would like to thank the the sponsors for their support. This work was funded in part by the National Science Foundation (NSF) under Grant CNS-1642873, and the Defense University Research Instrumentation Program (DURIP) contract numbers W911NF-14-1-0553/0554 through the Army Research Office. 
\balance
\bibliographystyle{IEEEtran}
\bibliography{references_globecom}
\end{document}